\documentclass[]{arXiv}
\usepackage{amsmath,amssymb,amsfonts}
\usepackage{algorithmic}
\usepackage{graphicx}
\usepackage{textcomp}
\usepackage{algorithm}
\usepackage{bm}
\usepackage{rotating}
\usepackage{multirow}
\usepackage{comment}
\usepackage{tabularx}
\usepackage{subcaption}
\usepackage{xcolor}
\usepackage{makecell}
\usepackage{array}

\title{Evolution of Lane-Changing Behavior in Mixed Traffic:\\A Quantum Game Theory Approach}

\author[1]{Sungyong Chung\orcidlink{0009-0001-3131-6011}}
\author[1]{Tina Radvand\orcidlink{0000-0002-4454-8567}}
\author[1]{Alireza Talebpour\orcidlink{0000-0002-5412-5592}\thanks{Corresponding author. Email: ataleb@illinois.edu}}

\affil[1]{Department of Civil and Environmental Engineering, University of Illinois Urbana-Champaign}

\date{}

\begin{document}
\maketitle

\begin{abstract}
As automated vehicles (AVs) enter mixed traffic, proactively anticipating the evolution of human driving behavior during critical interactions, such as lane changes, is essential. However, classical Evolutionary Game Theory (EGT) fails to capture the complexity of human decision-making during lane changes. Specifically, by strictly assuming independence between agents, classical models calibrated on empirical payoffs predict a convergence to unrealistic full cooperation, contradicting the stable 42\% cooperation rate observed in real-world data. To resolve this discrepancy, this study introduces a Quantum Game Theory (QGT) framework. We analyze 7,636 lane-changing interactions from the Waymo Open Motion Dataset (WOMD) to derive empirical payoff matrices via a Quantal Response Equilibrium (QRE) model. Utilizing the Marinatto-Weber (MW) quantization scheme, we introduce an entanglement parameter to mathematically embed latent correlations directly into the payoff structure of a single interaction. Our results identify a human entanglement parameter of $|b|^2_{HDV} \approx 0.52$ that accurately reproduces the observed mixed equilibrium. Furthermore, simulations of three AV deployment strategies (classical, entangled, and inverted) reveal that human adaptation depends critically on the underlying AV algorithm: while cooperative classical AVs maximize system-wide cooperation at high market penetration rates, defective inverted AVs paradoxically yield higher overall cooperation at low penetration rates by prompting more cooperative behaviors from human drivers. Consequently, rather than waiting for large scale deployment to observe these effects, stakeholders can utilize this framework to simulate repeated interactions and proactively anticipate how human driver behavior will evolve in response to specific AV software designs.
\end{abstract}

\keywords{Automated Vehicle, Cooperation, Evolutionary Game Theory, Lane-Changing, Quantum Game Theory}

\section{Introduction}
\label{sec:introduction}

Automated vehicles (AVs) are gradually entering road networks long dominated by human-driven vehicles (HDVs). This creates a transitional era of mixed traffic in which AVs and HDVs must operate side by side. Among the many driving tasks required in this mixed traffic environment, lane changes are especially critical to study because they require anticipating both available gaps and the intentions and reactions of nearby vehicles. Although substantial work has examined the control and planning algorithms of AVs \cite{An2023Bezier, An2023Platooning, Rahmati2021LeftTurn, Chung2024GapSetting}, interactive decision-making between AVs and HDVs during lane changes remains largely unexplored \cite{Chung2025}. This gap is important because it is still unclear whether opportunistic human drivers will exploit AVs or whether the presence of AVs will promote more cooperative behavior in traffic.

In our previous work \cite{Chung2025}, we analyzed real-world trajectories from the Waymo Open Motion Dataset (WOMD) \cite{Ettinger2021} to characterize lane-changing interactions. We mapped lane-changing events to a game-theoretic framework involving active (lane-changing) and passive (target-lane lag) vehicles. The observed behaviors were clustered into cooperative and defective, and corresponding payoffs were then estimated for each strategic type. The results indicated that approximately 42\% of human drivers exhibited cooperative behavior during lane-changing. However, in this study, when these empirically derived payoffs are applied in a classical Evolutionary Game Theory (EGT) simulation of a 100\% HDV population, the system rapidly converges toward near-full cooperation. This reveals a key inconsistency: the estimated payoffs imply that the current mixed state should be unstable and evolve toward full cooperation, yet actual observed traffic behavior persists in a stable, sub-optimal mixed equilibrium.

This discrepancy highlights a fundamental limitation of classical EGT. While EGT effectively models how a population learns and updates strategies over time based on past payoffs, it fundamentally assumes that the execution of each individual encounter is independent. In the classical framework, agents in a single-shot game act without any latent correlation during the interaction itself. In reality, human drivers bring accumulated experience, such as established social norms and driving culture, into every interaction. These latent factors inherently correlate their choices, even in a single-shot encounter with a stranger. 

By introducing an entanglement parameter, Quantum Game Theory (QGT) mathematically embeds this shared experience directly into the formulation of a single game, capturing the correlations that classical independence assumptions cannot. Specifically, QGT extends classical game theory by allowing for entanglement between strategies of players. Here, entanglement is not interpreted as a physical quantum phenomenon, but rather as a mathematical formalism that models correlations and interdependence between decision-makers \cite{Khan2018}.

In this study, we use the QGT framework to model the evolution of lane-changing behavior. We specifically employ the Marinatto-Weber (MW) quantization scheme \cite{Marinatto2000}, which enables the existence of evolutionarily stable strategies in mixed equilibria where classical theory fails \cite{Iqbal2001}. Within this framework, we treat entanglement as a necessary mathematical tool to stabilize the system at the observed 42\% cooperation rate in 100\% HDV traffic. We then extend this framework to mixed traffic by introducing three types of AVs: (i) classical AVs, which behave as fully rational independent agents with zero entanglement, (ii) entangled AVs, which utilize vehicle-to-vehicle (V2V) communication to achieve strong coordination, and (iii) inverted AVs, which adopt strategies opposite to those of classical AVs under an inverted reward structure. We perform simulations across a range of AV market penetration rates (MPRs) and obtain insight into AV deployment strategies that guide mixed traffic toward more cooperative behavior.

\section{Literature Review}
Game theory has been widely used to model decision making in vehicle interactions, particularly for lane changes where drivers must anticipate the responses of others \cite{Kita1999, Kita2002, Talebpour2015}. Most game-theoretic lane-changing studies have focused on modeling the lane-changing decision itself, with few studies examining the cooperative or defective behaviors that emerge during these interactions. For instance, \citep{Tanimoto2014} defined cooperation as remaining in the current lane and defection as changing lanes to move ahead, and showed through numerical experiments that lane-changing interactions can exhibit Prisoner's Dilemma dynamics at moderate traffic densities. More recently, our earlier work \cite{Chung2025} used real-world trajectory data to analyze lane-changing interactions and provided the first empirical evidence that (i) AVs exhibit higher cooperation rates than HDVs in both active and passive roles, (ii) social dilemmas appear in both AV–HDV and HDV–HDV interactions, and (iii) classical EGT predicts increased cooperation for both vehicle types as AVs enter the traffic stream.

A key limitation of classical EGT, however, is its assumption that players randomize their strategies independently, which forces the joint probability distribution to be factorizable \cite{Khan2018}. This assumption confines evolutionary dynamics to a narrow subset of possible outcomes and fails to capture correlated behavior that enables the resolution of social dilemmas in complex settings \cite{Grabbe2005}. QGT provides an extension of the classical framework that can address these limitations. In the following subsections, we outline the fundamental principles of QGT (Section \ref{sec:quantum_game_theory}) and introduce the two principal quantization schemes: the Eisert–Wilkens–Lewenstein (EWL) scheme \cite{Eisert2000} and the MW scheme \cite{Marinatto2000} (Section \ref{sec:quantization_scheme}).

\subsection{Quantum Game Theory}\label{sec:quantum_game_theory}
QGT was developed to extend classical games into the setting of quantum information processing \cite{meyer1999}. It replaces classical probabilities with quantum amplitudes and introduces entanglement as a resource that creates correlation between players. In the context of social interaction, this formalism is not viewed as a physical effect at the subatomic level. Instead, it is reinterpreted through the lens of social equilibrium \cite{Debreu1952}, where strategy sets of players are mutually constrained by factors such as social norms or trust. Entanglement functions as a form of mediated communication that enables correlated equilibria \cite{Aumann1974}. However, unlike classical correlated equilibria that require an external physical signaling mechanism to coordinate players, quantum entanglement naturally embeds latent correlations directly into the state space through entanglement. This allows the system to stabilize outcomes that are inaccessible to independent classical agents, without the need for an external observer. As \citep{Khan2018} noted, QGT provide a game-theoretic framework for designing mechanisms that achieve desirable outcomes under multiple constraints, effectively using the quantum structure to overcome strict assumptions of independence in classical models.

\subsection{Quantization Schemes}\label{sec:quantization_scheme}
\begin{figure}[tb!]
    \centering
    \includegraphics[width=0.45\columnwidth]{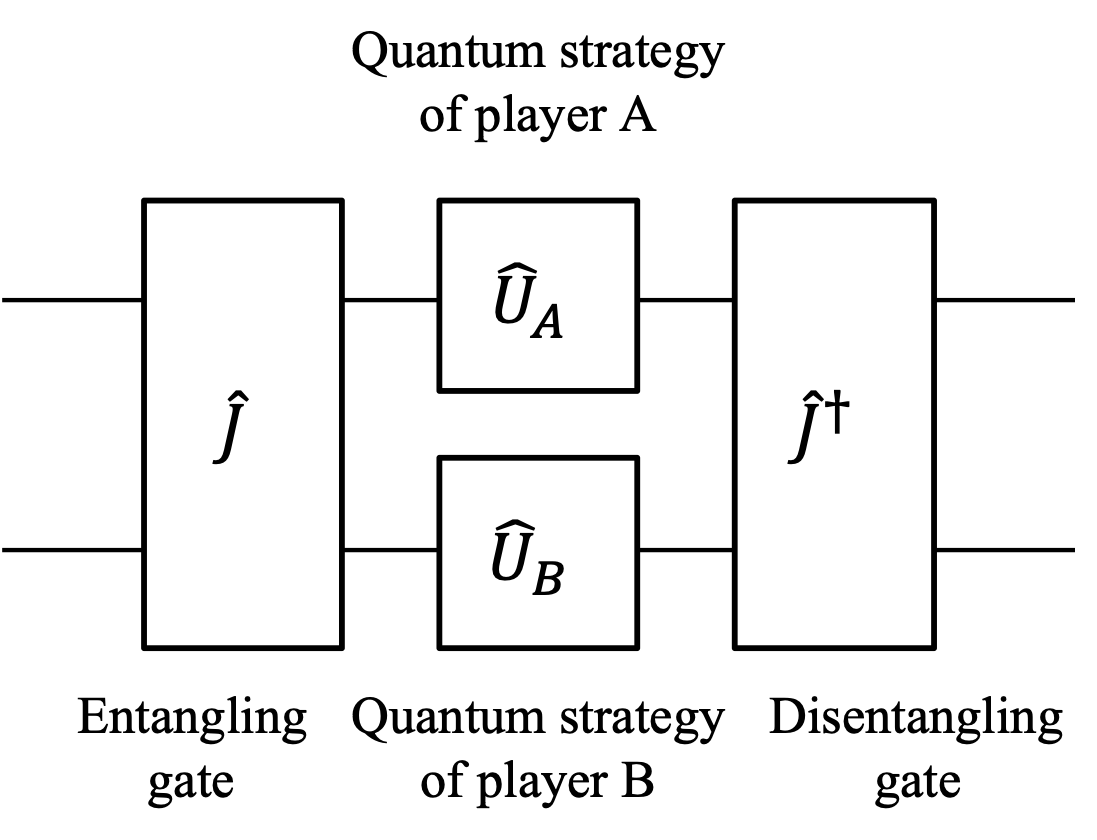}
    \caption{Quantum circuit diagram for the EWL game quantization scheme \cite{Eisert2000}. This scheme uses an entangling gate $\hat{J}$ and a disentangling gate $\hat{J}^{\dagger}$ surrounding the players' unitary strategies. Diagram adapted from \citep{Khan2018}.}
    \label{fig:scheme_ewl}
\end{figure}

\begin{figure}[tb!]
    \centering
    \includegraphics[width=0.45\columnwidth]{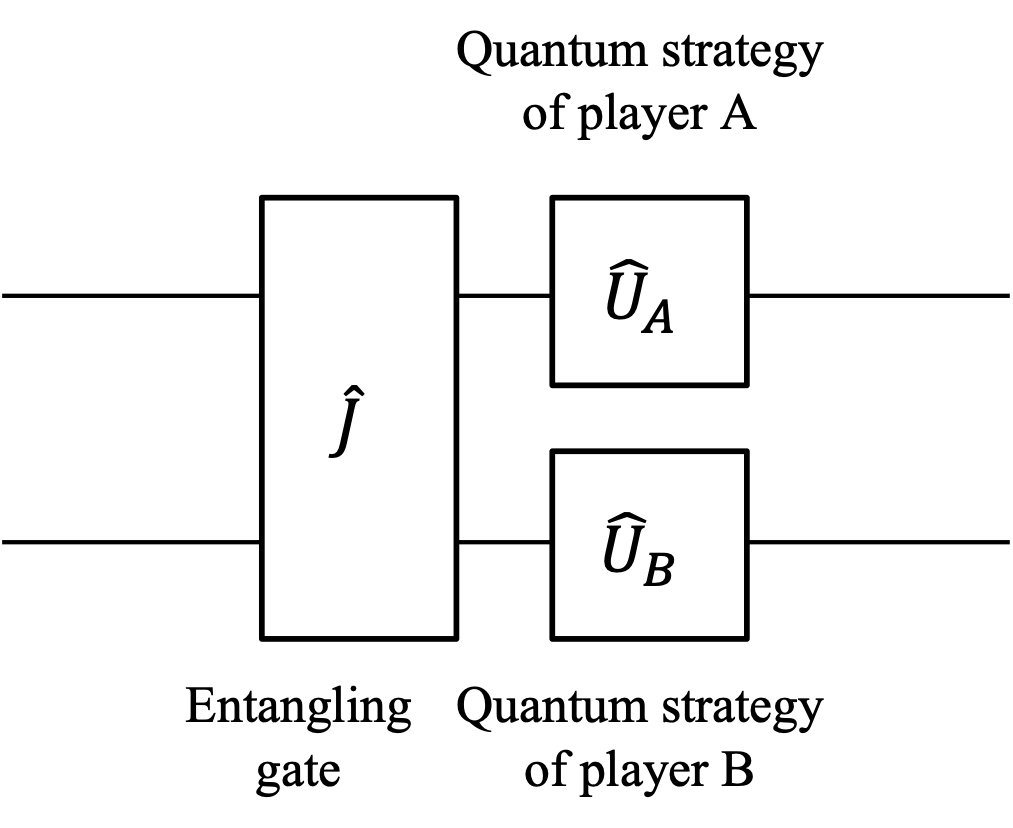}
    \caption{Quantum circuit diagram for the MW game quantization scheme \cite{Marinatto2000}. This scheme removes the disentangling gate, and players apply classical strategies (such as $\hat{I}$ or $\hat{\sigma}_{x}$) directly to the entangled state. Diagram adapted from \citep{Khan2018}.}
    \label{fig:scheme_mw}
\end{figure}

Two principal approaches have been used to quantize classical games: the EWL scheme \cite{Eisert2000} and the MW scheme \cite{Marinatto2000}. The structural differences between these schemes are summarized in Figures \ref{fig:scheme_ewl}-\ref{fig:scheme_mw}.

The EWL scheme (Figure \ref{fig:scheme_ewl}) formulates the game as a quantum circuit in which players manipulate qubits with unitary operators drawn from a continuous strategy space. The scheme includes an entangling operator $\hat{J}$ applied prior to the strategies and a disentangling operator $\hat{J}^\dagger$ applied just before measurement. In the setting of the Prisoner's Dilemma, \citep{Eisert2000} showed that this structure allows players to move beyond the classical Nash equilibrium \cite{nash1950equilibrium} of mutual defection. By adopting a particular quantum strategy $\hat{Q}$, both players can reach a unique, Pareto-optimal Nash equilibrium corresponding to full mutual cooperation.

The MW scheme (Figure \ref{fig:scheme_mw}) takes a fundamentally different approach by omitting the disentangling gate. To build an intuition for how the MW scheme mathematically operates, it is instructive to examine its mechanics. Departing from classical games where players make independent decisions from a separable state, this model initiates with an entangled state that represents the inherent latent correlations between agents. The initial state is defined in the computational basis as:
\begin{equation}
\lvert \psi_{in} \rangle = \hat{J} \lvert 00 \rangle = a \lvert 00 \rangle + b \lvert 11 \rangle,
\label{eq:initial_state}
\end{equation}
where the coefficients satisfy the normalization constraint $|a|^2 + |b|^2 = 1$. The parameter $|b|^2 \in [0, 1]$ governs the degree of entanglement.
This elegant mechanism of blending payoffs through entanglement mathematically rewrites the strategic landscape of classical game theory. This is demonstrated in two foundational base cases:

 \subsubsection{The Battle of the Sexes} 
In the classical Battle of the Sexes, two players (Alice and Bob) wish to coordinate their actions by choosing between two strategies: going to the Opera ($O$) or watching TV ($T$). While they both desire coordination, they prefer different outcomes. The game is defined by the following normal-form payoff bimatrix:
\begin{equation}
\bordermatrix{
  & O & T \cr
O & (\alpha, \beta) & (\gamma, \gamma) \cr
T & (\gamma, \gamma) & (\beta, \alpha) \cr
}
\end{equation}
where the first value in each cell represents the payoff to Alice, the second to Bob, and the condition $\alpha > \beta > \gamma$ holds. Classically, this yields a frustrating dilemma: there are two unequal pure-strategy Nash equilibria, $(O, O)$ and $(T, T)$, where one player wins their preferred choice while the other concedes. Furthermore, the game possesses one mixed-strategy equilibrium that yields a sub-optimal expected payoff for both players.

Marinatto and Weber \cite{Marinatto2000} demonstrated that the MW scheme resolves this coordination failure by reformulating the game using quantum operators. In this framework, payoffs are not directly assigned to the strategies chosen by the players. Instead, they are assigned to the final state of the system upon measurement. Consequently, the payoff matrix is redefined in the computational basis of the state space:
\begin{equation}
\bordermatrix{
  & |0\rangle & |1\rangle  \cr
|0\rangle  & (\alpha, \beta) & (\gamma, \gamma) \cr
|1\rangle   & (\gamma, \gamma) & (\beta, \alpha) \cr
}
\end{equation}
This payoff structure is formally encapsulated by the payoff operators ($\hat{\$}$) for Alice and Bob:
\begin{align}
    \hat{\$}_A &= \alpha|00\rangle\langle 00| + \gamma|01\rangle\langle 01| + \gamma|10\rangle\langle 10| + \beta|11\rangle\langle 11|, \label{eq:payoff_operator_A} \\
    \hat{\$}_B &= \beta|00\rangle\langle 00| + \gamma|01\rangle\langle 01| + \gamma|10\rangle\langle 10| + \alpha|11\rangle\langle 11|. \label{eq:payoff_operator_B}
\end{align}
Players determine the final state, $|\psi_{fin}\rangle$, by applying their chosen strategy, either the identity operator ($\hat{I}$) or the Pauli-X operator ($\hat{\sigma}_x$), to their respective qubits in a shared initial state defined as $|\psi_{in}\rangle = a|00\rangle + b|11\rangle$, where the first and second qubits correspond to Alice and Bob, respectively. In practical terms, the $\hat{I}$ operator leaves a player's qubit unchanged, while the $\hat{\sigma}_x$ operator acts as a quantum bit-flip, inverting the state between $|0\rangle$ and $|1\rangle$. The expected payoff is calculated based on the final state of the system by applying their respective payoff operators: $\langle\psi_{fin}|\hat{\$}_A|\psi_{fin}\rangle$ and $\langle\psi_{fin}|\hat{\$}_B|\psi_{fin}\rangle$. 

The parameter $|b|^2$ serves as the bridge between classical and quantum game theory. If $|b|^2 = 0$, the initial state is the separable state $|00\rangle$. In this classical limit, the strategy $\hat{I}$ (going to the Opera) maps directly to the basis state $|0\rangle$, and $\hat{\sigma}_x$ (watching TV) maps directly to $|1\rangle$. This perfectly recovers the classical game and its inherent dilemmas. For instance, if both Alice and Bob apply $\hat{I}$ (representing a mutual decision to go to the Opera), the final state remains $|00\rangle$, yielding the classical payoff $(\alpha, \beta)$.

However, as the entanglement parameter $|b|^2$ increases, the mapping between strategy and payoff shifts. If the game starts in a maximally entangled state ($|a|^2 = |b|^2 = 0.5$), and both players choose go to the Opera (applying $\hat{I} \otimes \hat{I}$), the final state preserves the initial superposition of $|00\rangle$ and $|11\rangle$. Consequently, the expected payoff of the mutual decision to go to the Opera is no longer $(\alpha, \beta)$, but rather a weighted average of $\alpha$ and $\beta$. This fundamental change in the payoff structure forces a unique, symmetric Nash equilibrium where both players obtain an equal expected payoff of $(\alpha+\beta)/2$, guiding both agents to a fair, optimal outcome without requiring direct communication or an external observer. 

To clearly distinguish this quantum mechanism from classical mixed strategies, consider a scenario where both players attempt to randomize their choices to achieve fairness. In a classical game, if Alice and Bob independently choose to play $O$ (represented by $\hat{I}$) with a 50\% probability and $T$ (represented by $\hat{\sigma}_{x}$) with a 50\% probability, the independence assumption mathematically forces the final joint distribution to be exactly 25\% for each of the four possible states: $|00\rangle$, $|01\rangle$, $|10\rangle$, and $|11\rangle$. Classical agents acting independently can never achieve a joint state distribution of exactly 50\% $|00\rangle$ and 50\% $|11\rangle$. However, in the quantum framework, when both players apply the pure, deterministic strategy $\hat{I}$ to a maximally entangled initial state, the system naturally yields this perfect 50/50 correlation. This demonstrates how entanglement bypasses the classical independence assumption, embedding latent correlation directly into the payoff structure of a single interaction.

\subsubsection{The Prisoner's Dilemma}
In the classical prisoner's dilemma, two players choose between cooperation ($C$) and defection ($D$). The payoff structure is determined by the reward $R$, punishment $P$, temptation $T$, and sucker's payoff $S$, subject to the inequality $T > R > P > S$. The game is characterized by the following bimatrix:

\begin{equation}
\bordermatrix{
  & C & D \cr
C & (R, R) & (S, T) \cr
D & (T, S) & (P, P) \cr
}
\end{equation}

Classically, defection ($D$) strictly dominates cooperation ($C$). Regardless of the opponent's choice, a rational player always gains a higher payoff by defecting. Consequently, mutual defection $(D, D)$ is the unique Nash equilibrium and the only Evolutionarily Stable Strategy (ESS), leading to a sub-optimal outcome for both players.

Iqbal and Toor \cite{Iqbal2001} applied the MW quantization scheme to the prisoner's dilemma, defining the payoff operators in the computational basis as:
\begin{equation}
\bordermatrix{
  & |0\rangle & |1\rangle  \cr
|0\rangle  & (R, R) & (S, T) \cr
|1\rangle   & (T, S) & (P, P) \cr
}
\end{equation}
The payoff operators for Alice and Bob are formally written as:
\begin{align}
    \hat{\$}_A &= R|00\rangle\langle 00| + S|01\rangle\langle 01| + T|10\rangle\langle 10| + P|11\rangle\langle 11|, \\
    \hat{\$}_B &= R|00\rangle\langle 00| + T|01\rangle\langle 01| + S|10\rangle\langle 10| + P|11\rangle\langle 11|.
\end{align}
As in the previous case, players manipulate a shared entangled state $|\psi_{in}\rangle = a|00\rangle + b|11\rangle$ using the identity operator $\hat{I}$ (representing the strategy of cooperation) or the Pauli-X operator $\hat{\sigma}_x$ (representing the strategy of defection). In the classical limit where $|b|^2=0$, if both players choose to cooperate (applying $\hat{I} \otimes \hat{I}$), the final state remains strictly $|00\rangle$, yielding the classical payoff $(R, R)$.

Iqbal and Toor \cite{Iqbal2001} demonstrated that the introduction of entanglement dramatically alters the evolutionary dynamics of the game. Using the standard payoff values ($R=3, P=1, T=5, S=0$), they identified three distinct behavioral regimes determined solely by the magnitude of the entanglement parameter $|b|^2$. In the regime of low entanglement ($|b|^2 \le 1/3$), the game behaves qualitatively like the classical prisoner's dilemma, where the temptation to defect dominates; consequently, the strategy of defection ($\hat{\sigma}_x$) remains the unique ESS. Conversely, when entanglement is high ($|b|^2 \ge 2/3$), the strong correlation between players effectively inverts the dilemma, making the strategy of cooperation ($\hat{I}$) the unique ESS.

Crucially, a phase transition occurs in the intermediate range ($1/3 < |b|^2 < 2/3$). In this regime, neither pure cooperation nor pure defection is evolutionarily stable. Instead, the unique ESS is a mixed strategy, where players adopt cooperation ($\hat{I}$) with a specific probability $p^*$ and defection ($\hat{\sigma}_x$) with probability $1-p^*$. This third regime is of particular significance to mixed traffic modeling, as it provides a theoretical mechanism explaining why human drivers, who operate under implicit social norms (represented by $|b|^2$), settle into a stable equilibrium of partial cooperation rather than converging to the extremes of full cooperation or full defection predicted by classical models.

\section{Methodology}

To analyze the evolution of lane-changing behavior, we utilize a data-driven game-theoretic framework. This study builds upon the empirical foundation established in our previous work \cite{Chung2025}, extending it from a classical analysis into the quantum domain to resolve the discrepancies between theoretical predictions and observed reality.

\subsection{Empirical Foundation}\label{sec:empirical}

We utilize the WOMD to capture observed interactions in a mixed traffic of AVs and HDVs. Our analysis focuses on $7,636$ lane-changing events extracted from the WOMD, involving two distinct players: the active (lane-changing) vehicle and the passive (target-lane lag) vehicle.

We use a Quantal Response Equilibrium (QRE) framework \cite{McKelvey1995, Chung2025} to quantify the utilities of the behavior of active and passive vehicles from observed actions and to determine the behavioral strategy each vehicle is likely to select during a lane change maneuver, given the state immediately preceding the lane change. Unlike Nash equilibria, which assume perfect rationality, QRE models bounded rationality through probabilistic strategy choice.

First, we applied k-means clustering ($k=2$) to the behavioral features of active and passive vehicles. These features were computed directly from the trajectory data during the lane change. The resulting clusters were labeled as cooperative ($C$) or defective ($D$) based on an interpretation of their kinematic characteristics. For active vehicles, the cluster labeled as defective exhibited more aggressive lane-changing behavior, including shorter lane-changing times, larger speed gains and lane-crossing angle, increased variability in speed, yaw rate, and lateral acceleration, and higher peak lateral and longitudinal accelerations. For passive vehicles, the defective cluster was distinguished by larger speed gains and more extreme maximum and minimum accelerations during the lane-changing maneuver.

Second, we defined the lane-changing state vector for the observed event $i$ as $\mathbf{S}_i= \{{s}_1, {s}_2, \dots, {s}_{11}\}_i$, which consists of kinematic variables observed immediately prior to the lane change initiation. These variables capture the dynamics of the active, passive, and the target-lane lead vehicles just before the lane change, and are summarized in Table~\ref{tab:lane_changing_variables}.

\begin{table}[tb!]
    \centering
    \caption{Lane-changing State Variables (Adapted from \cite{Chung2025}).}
    \begin{tabular}{ll}
        \hline
        \textbf{Variable} & \textbf{Description} \\
        \hline
        $s_1$ & Active vehicle average speed before LC ($m/s$) \\
        $s_2$ & Active vehicle standard deviation of speed before LC ($m/s$) \\
        $s_3$ & Active Vehicle average acceleration before LC ($m/s^2$) \\
        $s_4$ & Average lead gap before LC ($s$) \\
        $s_5$ & Lead vehicle relative average speed before LC ($m/s$) \\
        $s_6$ & Lead vehicle standard deviation of speed before LC ($m/s$) \\
        $s_7$ & Lead vehicle average acceleration before LC ($m/s^2$) \\
        $s_8$ & Average lag gap before LC ($s$) \\
        $s_9$ & Passive vehicle relative average speed before LC ($m/s$) \\
        $s_{10}$ & Passive vehicle standard deviation of speed before LC ($m/s$) \\
        $s_{11}$ & Passive vehicle average acceleration before LC ($m/s^2$) \\
        \hline
    \end{tabular}
    \label{tab:lane_changing_variables}
\end{table}

Next, to estimate the behavior selection (cooperative or defective) of active and passive vehicles during the lane change based on the conditions right before it (which is represented by the state vector $\mathbf{S}_i$), we model the utilities associated with cooperative ($C$) and defective ($D$) strategies. The utility $U_{r, t, o}$ for a vehicle with role $r \in \{\text{active, passive}\}$ and interaction type $ t \in \{\text{AV (vs. HDV)}$, $\text{HDV (vs. AV)}$, $\text{HDV (vs. HDV)}$, $\text{AV (vs. AV)}\}$, for the outcome $o \in \{CC, CD, DC, DD\}$ is modeled as a linear function of the state variables. In the outcome $o$, the first and second letters denote the strategies of the active and passive vehicles, respectively. Formally, this linear utility specification is given by:

\begin{equation}
U_{r, t, o}(\textbf{S}_i) = \bm{\beta}_{r,t,o} \cdot \tilde{\mathbf{S}_i}, 
\label{eq:utility}
\end{equation}
where \( \tilde{\mathbf{S}_i} = \{1, \tilde{s}_1, \tilde{s}_2, \dots, \tilde{s}_{11}\}_i \) is the pre-processed state vector for event $i$, constructed by first standardizing the raw state vector $\mathbf{S}_i$ to zero mean and unit variance, followed by the inclusion of a unity element at the beginning to account for the model intercept. $\bm{\beta}_{r,t,o}$ denotes the vector of coefficients estimated in \citep{Chung2025}.

Finally, we map the estimated utilities into the standard payoff matrix required for the game theory-based simulation. Agian, the reward $R$ corresponds to mutual cooperation, the sucker's payoff $S$ corresponds to cooperating while the opponent defects, the temptation $T$ corresponds to defecting while the opponent cooperates, and punishment $P$ corresponds to mutual defection. For any observed state $\mathbf{S}_i$, the empirical payoffs are calculated as:

\begin{equation}
\label{eq:R}
R_{r, t}(\mathbf{S}_i) = U_{r, t, CC}(\mathbf{S}_i),
\end{equation}
\begin{equation}
S_{r, t}(\mathbf{S}_i) = 
\begin{cases} 
U_{r, t, CD}(\mathbf{S}_i) & \text{if } r = \text{active} \\
U_{r, t, DC}(\mathbf{S}_i) & \text{if } r = \text{passive} 
\end{cases}
\end{equation}
\begin{equation}
T_{r, t}(\mathbf{S}_i) = 
\begin{cases} 
U_{r, t, DC}(\mathbf{S}_i) & \text{if } r = \text{active} \\
U_{r, t, CD}(\mathbf{S}_i) & \text{if } r = \text{passive}
\end{cases}
\end{equation}
\begin{equation}
P_{r, t}(\mathbf{S}_i) = U_{r, t, DD}(\mathbf{S}_i) = 0.
\end{equation}
The baseline utility for mutual defection is normalized to zero during estimation.

It is important to note that in our previous work \cite{Chung2025}, we found that approximately 4\% and 11\% of the 7,636 observed lane-changing events for active and passive vehicles, respectively, were identified as social dilemmas. Among these, the majority were classified as either Stag Hunt or Prisoner's Dilemma games, while the Chicken game was rarely observed. Because the empirical payoffs are calculated individually for each unique lane-changing state $S_i$, the dataset inherently encapsulates a diverse array of game classes across the mixed traffic environment.

\subsection{Quantum Game Theory}\label{sec:qgt}

The payoffs derived in Section \ref{sec:empirical} via the QRE model quantify the baseline utility of a strategy under the classical assumption of independent decision-making. While this approach accurately evaluates the isolated utility of strategies, it fundamentally fails to capture the interdependent nature of real-world traffic during repeated interactions, where decisions are influenced by latent correlations such as social norms and driving culture. In other words, the limitation of the classical evolutionary framework does not lie in how it evaluates the isolated utility of a driving outcome, but rather in its assumption that agents execute these choices independently during population-level interactions.

To address this, we employ the MW quantization scheme \cite{Marinatto2000}, which utilizes these classically derived payoffs as the base game matrix but relaxes the assumption of independence during the interaction. By applying quantum operators, the entanglement parameter $|b|^2$ serves as a mathematical bridge, allowing us to embed these correlations
directly into the payoff structure of a single interaction.

In the MW scheme (Figure \ref{fig:scheme_mw}), the interaction between two players, $A$ and $B$, representing active and passive vehicles, respectively, is modeled. The game is formulated in a Hilbert space $\mathcal{H} = \mathcal{H}_A \otimes \mathcal{H}_B$, which constitutes the tensor product of the individual state spaces $\mathcal{H}_A$ and $\mathcal{H}_B$. Each player's subspace is spanned by the computational basis $\{\lvert 0\rangle, \lvert 1\rangle\}$. Deviating from classical games that typically utilize separable initial states, the MW approach employs an entangled initial state:
\begin{equation}
\lvert \psi_{in} \rangle = a \lvert 00 \rangle + b \lvert 11 \rangle,
\end{equation}
where $\lvert b\rvert^{2} \in [0, 1]$. The initial state reaches maximal entanglement when $\lvert b\rvert^{2} = 0.5$.

Strategies are chosen by the players through the application of unitary operators to the initial state. Within the MW framework, the set of available strategies is restricted to the identity operator $\hat{I}$ and the Pauli X operator $\hat{\sigma}_{x}$. In alignment with the standard interpretation established by \citep{Marinatto2000} and \citep{Iqbal2001}, strategy $\hat{I}$ represents cooperation, and strategy $\hat{\sigma}_{x}$ represents defection.

If the active vehicle applies strategy $\hat{U}_{active} \in \{\hat{I}, \hat{\sigma_x}\}$ and the passive vehicle applies $\hat{U}_{passive} \in \{\hat{I}, \hat{\sigma_x}\}$, the final state becomes:
\begin{equation}
\lvert \psi_{fin} \rangle = (\hat{U}_{active} \otimes \hat{U}_{passive}) \lvert \psi_{in} \rangle.
\end{equation}

The expected payoff for each player is given by the expectation value of the corresponding payoff operator, $\langle\psi_{fin}|\hat{\$}_i|\psi_{fin}\rangle$, where the payoff operators $\hat{\$}_A$ and $\hat{\$}_B$ are defined in Equations \ref{eq:payoff_operator_A} and \ref{eq:payoff_operator_B}. 

Recall that if both players choose the strategy of cooperation (applying \(\hat{I} \otimes \hat{I}\)), the state remains entangled as \(a \lvert 00 \rangle + b \lvert 11 \rangle\). Consequently, the expected payoff for mutual cooperation is a weighted average of the classical Reward ($R$, associated with $|00\rangle$) and Punishment ($P$, associated with $|11\rangle$). By generalizing this logic to all strategy combinations, we derive the quantum expected payoffs ($R^q$, $S^q$, $T^q$, $P^q$) for a vehicle with role $r$ and interaction type $t$ in state $\mathbf{S}_i$:
\begin{equation} \label{eq:R}
R^q_{r,t}(\mathbf{S}_i) = R_{r,t}(\mathbf{S}_i)(1 - \lvert b\rvert^{2}) + P_{r,t}(\mathbf{S}_i)\lvert b\rvert^{2},
\end{equation}
\begin{equation} \label{eq:S}
S^q_{r,t}(\mathbf{S}_i) = S_{r,t}(\mathbf{S}_i)(1 - \lvert b\rvert^{2}) + T_{r,t}(\mathbf{S}_i)\lvert b\rvert^{2},
\end{equation}
\begin{equation} \label{eq:T}
T^q_{r,t}(\mathbf{S}_i) = T_{r,t}(\mathbf{S}_i)(1 - \lvert b\rvert^{2}) + S_{r,t}(\mathbf{S}_i)\lvert b\rvert^{2},
\end{equation}
\begin{equation} \label{eq:P}
P^q_{r,t}(\mathbf{S}_i) = P_{r,t}(\mathbf{S}_i)(1 - \lvert b\rvert^{2}) + R_{r,t}(\mathbf{S}_i)\lvert b\rvert^{2}.
\end{equation}

\noindent Here, $R^q_{r,t}(\mathbf{S}_i)$ represents the expected payoff under mutual cooperation (\(\hat{I} \otimes \hat{I}\)), while $S^q_{r,t}(\mathbf{S}_i)$ denotes the return when a player cooperates (\(\hat{I}\)) against a defecting opponent (\(\hat{\sigma}_x\)). Conversely, $T^q_{r,t}(\mathbf{S}_i)$ signifies the payoff for defecting (\(\hat{\sigma}_x\)) against a cooperative opponent (\(\hat{I}\)), and $P^q_{r,t}(\mathbf{S}_i)$ corresponds to the outcome when both participants choose to defect (\(\hat{\sigma}_x \otimes \hat{\sigma}_x\)).

This formulation demonstrates how the entanglement parameter \(\lvert b\rvert^{2}\) reshapes the underlying game. As shown in \ref{eq:R}-\ref{eq:P}, when \(\lvert b\rvert^{2}=0\), the quantum payoffs reduce exactly to the classical payoffs ($R^q=R$, $S^q=S$, $T^q=T$, and $P^q=P$), thereby recovering the standard EGT dynamics.

\subsection{Simulation Framework and Calibration}
We propose a simulation framework that integrates the MW scheme into an evolutionary setting. In the mixed traffic environment, we assign distinct entanglement parameters to each vehicle type, denoted as $\lvert b\rvert^2_{HDV}$ and $\lvert b\rvert^2_{AV}$. To align the model with empirical observations, we first calibrate the HDV entanglement level by sweeping $\lvert b\rvert^{2}_{HDV} \in [0, 1]$ in a 100\% HDV environment. The objective is to calibrate the value of \(\lvert b\rvert^{2}_{HDV}\) that yields an evolutionary equilibrium consistent with the empirically observed cooperation rate specific to HDV-HDV interactions, which is approximately 42\% \cite{Chung2025}. The resulting value is then adopted as the baseline representation of human behavior in mixed traffic settings.

With the HDV baseline established, we introduce AVs into the environment. Due to the lack of empirical observations for AV-AV interactions, we explore the behavioral design space by defining three distinct AV profiles: (i) classical AVs, in which AVs behave as independent rational agents without coordination and thus have $\lvert b \rvert_{AV}^{2} = 0.0$; (ii) entangled AVs, in which AVs are perfectly coordinated through V2V communication at a state of maximum entanglement and thus have $\lvert b \rvert_{AV}^{2} = 0.5$; and (iii) inverted AVs, which adopt strategies opposite to those of classical AVs under an inverted reward structure and thus have $\lvert b \rvert_{AV}^{2} = 1.0$. For interactions involving different vehicle types (for instance, an active AV interacting with a passive HDV), we assume that the effective entanglement level is the arithmetic mean of the two internal parameters:
\begin{equation}
\lvert b\rvert^{2}_{Mixed} = \frac{\lvert b\rvert^{2}_{AV} + \lvert b\rvert^{2}_{HDV}}{2}.
\end{equation}
We note that although we refer to these three AV types as classical, entangled, and inverted, the terminology is conceptual rather than definitive.

Following the evolutionary game formulations pioneered by \citep{nowak1992evolutionary}, and adapting the simulation framework developed in \citep{Chung2025}, the evolution of cooperative behavior is modeled on a \(20 \times 20\) lattice grid. At each time step \(t\), agents interact with neighbors within a specified radius and accumulate payoffs based on the quantum expected payoffs described in Section \ref{sec:qgt}. Strategies are updated using a pairwise comparison mechanism, where a vehicle adopts the strategy of a randomly selected neighbor of the same type. The probability of this transition is determined by the difference in their average payoffs and the noise parameter $K$. The complete procedure is provided in Algorithm \ref{alg:quantum_simulation}.

\begin{algorithm}[tb]
\caption{Quantum Evolutionary Game for Mixed Traffic}
\label{alg:quantum_simulation}
\begin{algorithmic}[1]
\REQUIRE Simulation parameters (grid size, AV MPRs, interaction neighbor size $d$, noise parameter $K$, social contact frequency $s$); Initial strategy/type distributions; Empirical payoffs $R_{r,t}(\mathbf{S}_i)$, $S_{r,t}(\mathbf{S}_i)$, $T_{r,t}(\mathbf{S}_i)$, $P_{r,t}(\mathbf{S}_i)$ from \citep{Chung2025}; Entanglement parameters $|b|^2_{HDV}$ and $|b|^2_{AV}$
\ENSURE Evolution of cooperative ratio for AVs and HDVs

\STATE \textbf{// Step 1: Initialization}
\STATE Construct a $20 \times 20$ grid of vehicles
\STATE Assign each grid cell a vehicle with a tuple $(\textit{Strategy}$, $\textit{Type})$ based on MPR and priors
\STATE Compute quantum expected payoffs $R^q_{r,t}(\mathbf{S}_i)$, $S^q_{r,t}(\mathbf{S}_i)$, $T^q_{r,t}(\mathbf{S}_i)$, $P^q_{r,t}(\mathbf{S}_i)$ for all states $\mathbf{S}_i$ and interaction pairs

\STATE \textbf{// Step 2: Simulation Loop}
\FOR{timestep $t=1$ to $T_{max}$}
    \STATE Randomly select a lane-changing state $\mathbf{S}_i$
    \FOR{each vehicle $X$}
        \STATE Identify neighbors based on interaction neighbor size $d$
        \STATE Retrieve precomputed Quantum Payoffs for state $\mathbf{S}_i$ based on vehicle types
        \STATE Compute average payoff $E_X$
    \ENDFOR
    \FOR{each vehicle $X$}
        \STATE Select random neighbor $Y$ of the \textit{same} vehicle type
        \STATE Calculate update probability $W$:
        \STATE \hfil $W = \displaystyle \frac{1}{1 + \exp\left[-(E_Y - E_X)/K\right]}$
        \vspace{4pt}
        \STATE Update $X$'s strategy to $Y$'s strategy with prob. $W$
    \ENDFOR
    \IF{$t \pmod{100/s} == 0$}
        \STATE Randomly shuffle vehicle positions
    \ENDIF
    \STATE Record the proportion of cooperative AVs and HDVs
\ENDFOR
\end{algorithmic}
\end{algorithm}

\section{Results and Discussions}

In this section, we present the results of our quantum evolutionary game simulations. First, we validate the framework by calibrating the entanglement parameter for human drivers, demonstrating that the QGT resolves the discrepancy between classical predictions and observed reality. Second, we apply this calibrated model to analyze the impact of different AV deployment strategies on mixed traffic cooperation.

\subsection{Calibration of Entanglement Parameter}
To assess the necessity of the quantum approach, we compared the evolutionary dynamics of a 100\% HDV population under different entanglement regimes. We simulated the evolution of cooperative behaviors over 200 time steps using the empirical payoff tables derived from the $7,636$ observed lane-changing events. Unless otherwise noted, simulations in this section were conducted with an interaction neighborhood size of 2 and a noise parameter of $K=2$.

\begin{figure}[tb!]
    \centering
    \includegraphics[width=0.6\columnwidth]{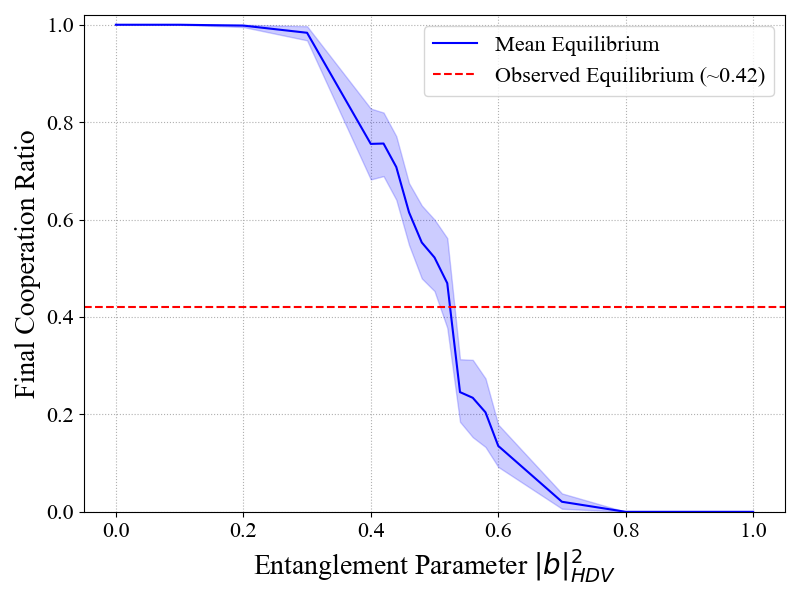}
    \caption{Calibration of the HDV-HDV entanglement parameter. The solid blue line represents the mean equilibrium cooperation rate across 20 simulations, with the shaded area indicating the 95\% confidence interval. The red dashed line represents the observed cooperation rate $\approx 0.42$. The intersection indicates a calibrated value of $\lvert b\rvert^{2}_{HDV} \approx 0.52$.}
    \label{fig:calibration}
\end{figure}

Figure \ref{fig:calibration} illustrates how the evolutionary outcome varies with the HDV entanglement parameter $\lvert b\rvert^{2}_{HDV}$. The x and y axes represent the entanglement parameter value and the final cooperation ratio of the population after 200 time steps, respectively, where the cooperation ratio is calculated as the proportion of agents in the grid that adopt strategy $\hat{I}$. The results indicate a transition from full cooperation to full defection as the entanglement parameter, $\lvert b\rvert^2_{HDV}$, increases from 0 to 1.

Importantly, under the classical model (\(\lvert b\rvert^{2}_{HDV} = 0\)), the predicted cooperation rate approaches 100\%, substantially exceeding empirical observations. Exploring the full range of entanglement values shows that $\lvert b\rvert^{2}_{HDV} \approx 0.52$ yields a stable evolutionary equilibrium consistent with the observed cooperation rate of 42\%. We therefore adopt $0.52$ as the HDV entanglement parameter.

\begin{figure}[tb!]
    \centering
    \includegraphics[width=0.6\columnwidth]{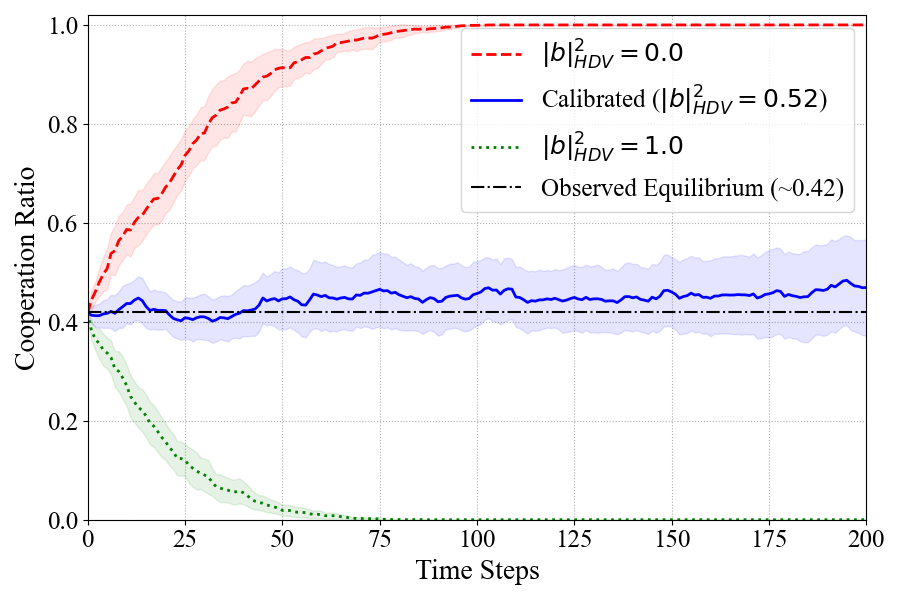}
    \caption{Evolution of cooperation in 100\% HDV traffic under different entanglement regimes. The classical model (red dashed, $|b|_{HDV}^2=0.0$) diverges from reality, while the calibrated quantum model (blue solid) accurately reproduces the observed mixed equilibrium.}
    \label{fig:calibration_evolution}
\end{figure}

Figure \ref{fig:calibration_evolution} compares the temporal evolution of cooperation ratio under three different cases. The first case (\(\lvert b\rvert^{2}_{HDV} = 0.0\)) represents independent rational agents, and the simulation quickly converges to full cooperation, failing to capture the complexity of real-world interactions. The second case (\(\lvert b\rvert^{2}_{HDV} = 1.0\)) corresponds to an initial state of $|11\rangle$, which in this dataset leads to a collapse of cooperation. Finally, the calibrated case (\(\lvert b\rvert^{2}_{HDV} \approx 0.52\)) successfully converges to and maintains the stable mixed equilibrium observed in the WOMD.

This result confirms that human lane-changing behavior is governed by latent correlations, which are analogous to the maximum level of entanglement and prevent the system from reaching the theoretically optimal state of full cooperation.

\subsection{Evolution of Cooperation in Mixed Traffic}\label{sec:evol_result}
In this section, we analyze how the introduction of AVs influences lane-changing behavior for both AVs and HDVs using the calibrated HDV entanglement parameter ($\lvert b\rvert^{2}_{\mathrm{HDV}} = 0.52$). Three deployment scenarios were analyzed: classical AVs (\(\lvert b\rvert^{2}_{AV} = 0.0\)), entangled AVs (\(\lvert b\rvert^{2}_{AV} = 0.5\)), and inverted AVs (\(\lvert b\rvert^{2}_{AV} = 1.0\)).

\begin{figure}[bt!]
    \centering
    \includegraphics[width=\columnwidth]{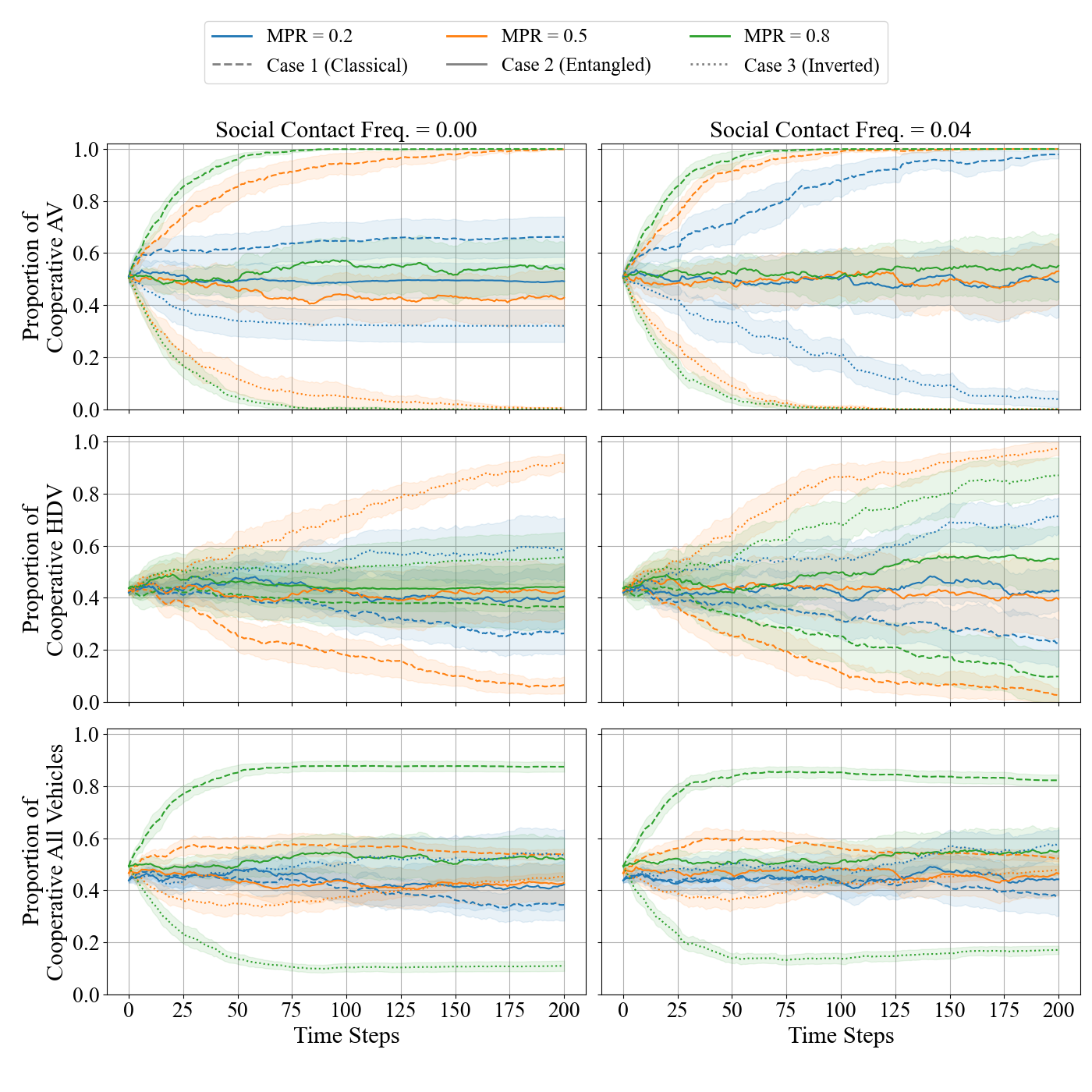}
    \caption{Comparison of evolutionary dynamics among classical AVs (dashed lines), entangled AVs (solid lines), and inverted AVs (dotted lines). The rows represent the cooperation ratio of AVs, HDVs, and all vehicles, respectively. The columns represent different social contact frequencies. A social contact frequency of 0.04 implies the grid is randomly shuffled four times over 100 steps, or once every 25 steps. The shaded regions depict the 95\% confidence intervals calculated from 20 simulation runs.}
    \label{fig:evolution_comparison}
\end{figure}

Figure \ref{fig:evolution_comparison} shows the evolution of cooperation ratios for AVs (top row), HDVs (middle row), and the overall system (bottom row) across different MPRs. We evaluate the three distinct AV behavioral profiles to understand their impact on mixed traffic dynamics.

In the classical AV scenario (Case 1), results demonstrate the potential risks of deploying purely rational, independent agents. Classical AVs (top row, dashed lines) consistently maintain high cooperation levels, but the HDV population (middle row, dashed lines) responds by reducing their own cooperation. Particularly in the well-mixed traffic at a MPR of 50\%, HDV cooperation declines substantially, suggesting that human drivers may exploit the cooperative AVs to maximize individual payoffs.

In the entangled AV scenario (Case 2), the evolutionary dynamics reveal a balanced system. Entangled AVs (top row, solid lines) stabilize at a symmetric mixed equilibrium of approximately 50\% cooperation across all MPRs. Because these AVs maintain a consistent, predictable level of partial cooperation through V2V coordination, they do not force human drivers to drastically alter their behavior. Consequently, the HDV population (middle row, solid lines) maintains a steady cooperation rate, preserving the natural traffic equilibrium without the exploitation observed in the classical case.

Conversely, the inverted AV scenario (Case 3) produces a notable asymmetry between vehicle types. As shown in the top row (dotted lines), inverted AVs rapidly converge toward a defective strategy due to their inverted reward structure, particularly under high MPR scenarios. Faced with this predictable, non-cooperating AV population, human drivers (middle row, dotted lines) are incentivized to exhibit a compensatory increase in cooperation to maximize their own payoffs. Consequently, although the inverted AVs themselves are less cooperative, their presence induces greater cooperation among human drivers.

Furthermore, a comparison between the left ($s=0.00$) and right ($s=0.04$) columns reveals that these evolutionary trends remain consistent regardless of the social contact frequency.

These results highlight a complex trade-off in AV deployment and behavioral design. At low penetration rates (MPR$= 0.2$), inverted AVs enhance overall system cooperation (bottom row, dotted lines) by effectively regulating HDV behavior. However, when AVs dominate the traffic (MPR$= 0.8$), the classical AV strategy becomes more advantageous for the system as a whole, since the high frequency of highly cooperative AV-AV interactions compensates for the reduced HDV cooperation (bottom row, dashed lines). Meanwhile, entangled AVs offer a robust middle ground across all penetration rates. By maintaining a stable, symmetric equilibrium, they prevent the human exploitation seen in the classical case while avoiding the defective dynamics of the inverted case.

While these three behavioral profiles serve as theoretical boundaries of AV algorithms, they demonstrate that different AV settings critically alter the evolution of human driving behaviors. Therefore, as AV deployment on roads increases, proactively evaluating how different AV software designs shape human adaptation will be essential for safe and efficient traffic integration.

\subsection{Sensitivity Analysis}

To evaluate the robustness of our simulation framework, we performed a sensitivity analysis on four key parameters: interaction neighborhood size, noise parameter ($K$), social contact frequency, and MPR. Unless otherwise specified, the default values were set as follows: interaction neighborhood size $d = 2$, noise parameter $K = 2$, social contact frequency $s = 0.02$, and MPR $= 0.5$.

\begin{figure}[tb!]
    \centering
    \includegraphics[width=0.7\columnwidth]{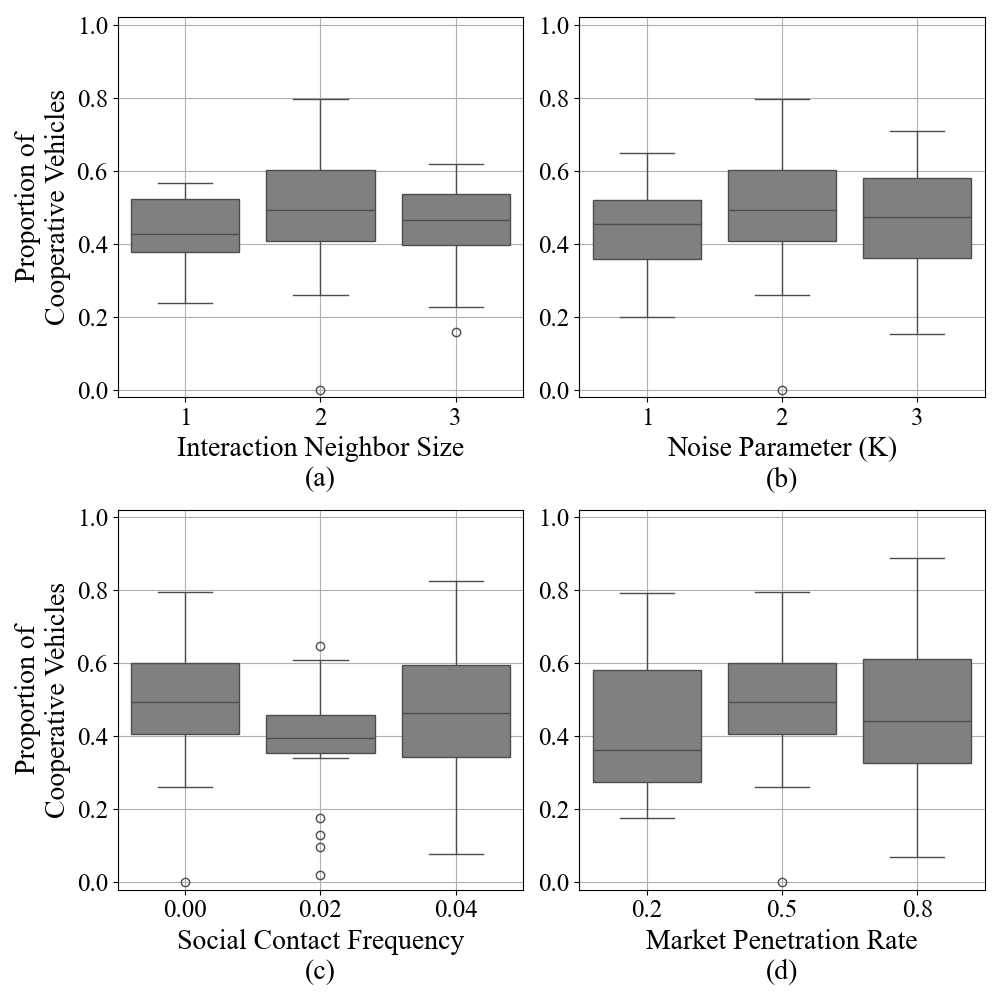}
    \caption{Sensitivity analysis for Case 1 (Classical AVs). Box plots show the distribution of the final proportion of cooperative vehicles after 200 time steps.}
    \label{fig:sensitivity_case1}
\end{figure}

\begin{figure}[tb!]
    \centering
    \includegraphics[width=0.7\columnwidth]{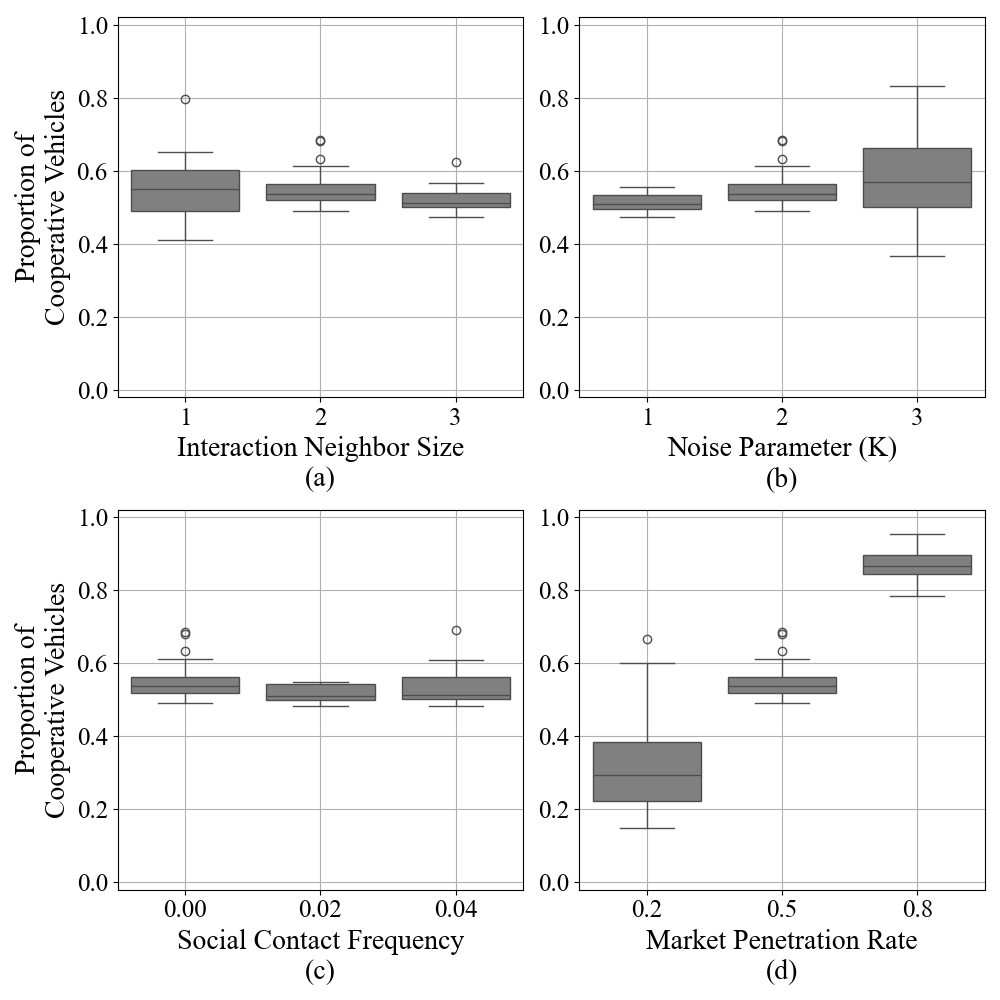}
    \caption{Sensitivity analysis for Case 2 (Entangled AVs). Box plots show the distribution of the final proportion of cooperative vehicles after 200 time steps.}
    \label{fig:sensitivity_case2}
\end{figure}

\begin{figure}[tb!]
    \centering
    \includegraphics[width=0.7\columnwidth]{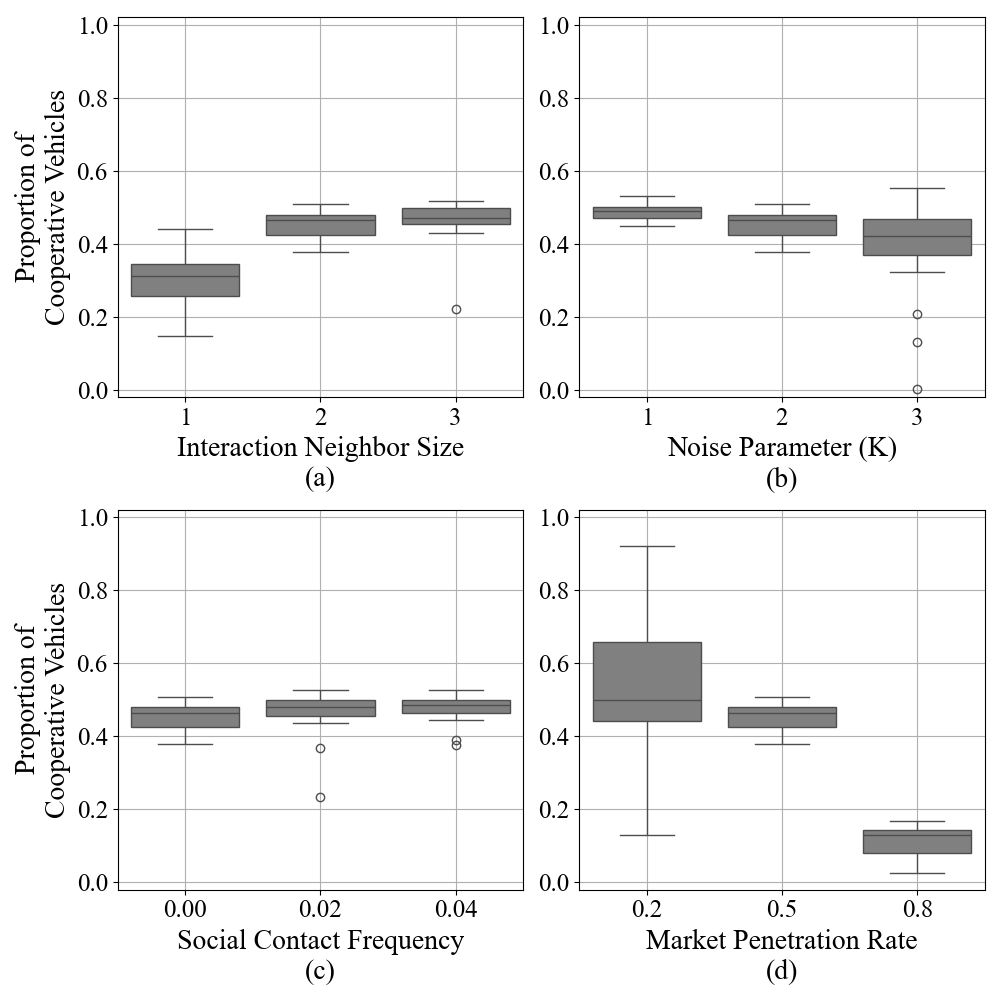}
    \caption{Sensitivity analysis for Case 3 (Inverted AVs). Box plots show the distribution of the final proportion of cooperative vehicles after 200 time steps.}
    \label{fig:sensitivity_case3}
\end{figure}

Figures \ref{fig:sensitivity_case1}, \ref{fig:sensitivity_case2}, and \ref{fig:sensitivity_case3} (a-c) present how the system responds to changes in hyperparameter settings across the three AV behavioral profiles. Interestingly, both Case 1 (classical AVs) and Case 3 (inverted AVs) exhibit similar sensitivities to local interaction and noise parameters. In both extreme cases, achieving a stable outcome with low variance requires sufficient local information and a decision process that is not overly dominated by randomness. Specifically, the results for interaction neighborhood sizes of 2 and 3, as well as noise parameters $K = 1$ and $K = 2$, produce tight, consistent distributions. However, when the neighborhood size is reduced to 1 or when the noise parameter is increased to $K = 3$, the variance grows noticeably. These patterns indicate that sparse local contact or excessive noise prevents the system from converging to a tight equilibrium. Accordingly, the default parameters (neighborhood size = 2, $K = 2$) were selected to ensure consistent evolutionary trends. 

In contrast, Case 2 (entangled AVs) displays a slightly different dynamic. While the median cooperation rate remains relatively stable near the 40--50\% range, the system exhibits substantial variance across all variations of neighborhood size, noise level, and social contact frequency. This suggests that the maximum entanglement introduces a wider spread of possible mixed-equilibrium states. Across all three cases, the results remain largely unchanged across different levels of social contact frequency, confirming that spatial shuffling has minimal impact.

Importantly, the system is most strongly affected by the MPR (Figures \ref{fig:sensitivity_case1}, \ref{fig:sensitivity_case2}, and \ref{fig:sensitivity_case3} (d)). In Case 1, overall cooperation rises dramatically with MPR, reflecting the increasing dominance of cooperative classical AVs. Conversely, in Case 3, overall cooperation collapses as MPR increases due to the growing influence of defective inverted AVs. The pronounced sensitivity to MPR relative to other hyperparameters highlights that the composition of vehicle types and their associated reward structures are the primary determinants of cooperative dynamics in mixed traffic.

\section{Conclusion}

This study introduced a QGT framework to address a fundamental limitation in classical traffic modeling: the rigid assumption of statistical independence during vehicle interactions. By treating lane-changing decisions as isolated events, classical EGT ignores the latent correlations, such as social norms and shared driving culture, that inherently constrain human behavior. Consequently, while classical models using empirical payoffs predict a convergence to unrealistic full cooperation, real-world data reveal that human drivers remain locked in a stable state of partial cooperation at around 42\%. To resolve this, we employed the MW quantization scheme to mathematically embed these latent correlations directly into the game structure via an entanglement parameter. Our simulations identified that an entanglement parameter of $|b|_{HDV}^{2}\approx0.52$ accurately reproduces the observed mixed equilibrium, providing the necessary mathematical bridge between theoretical predictions and observed reality.

Building upon this calibrated entanglement parameter, we evaluated how the introduction of AVs critically impacts the evolution of cooperative behaviors in mixed traffic. By evaluating three distinct AV behavioral profiles (classical, entangled, and inverted), we identified a complex trade-off in the AV deployment. We found that inverted AVs ($|b|_{AV}^{2}=1.0$), while tending toward defection, incentivize human cooperation. This makes them an effective strategy for achieving higher overall cooperation at low MPR. In contrast, classical AVs ($|b|_{AV}^{2}=0.0$), while tending toward cooperation, are exploited by human drivers at low penetration rates but maximize system-wide cooperation once they become the majority. Entangled AVs ($|b|_{AV}^{2}=0.5$) maintain a stable mixed equilibrium without inducing exploitative dynamics at any MPR.

Therefore, the primary contribution of this work is the development of a proactive behavior evaluation platform. As our simulations of three distinct AV profiles demonstrated, human behavior evolves differently depending on the underlying AV strategy, highlighting the critical need to anticipate these interactions. Manufacturers and regulators do not need to wait for large-scale AV deployment to observe long-term human adaptation. By mapping specific AV behavioral algorithms into this framework, stakeholders can simulate repeated interactions and anticipate how human drivers will evolve in response to their software. Future work will extend this game-theoretic modeling approach beyond passenger vehicles, utilizing macroscopic freight data to understand strategic interactions and flow evolution in commercial logistics.

\section*{Acknowledgments}
This work was supported by the National Science Foundation under Grant No. 2047937. 

\section*{Declaration of Generative AI and AI-assisted Technologies in the Writing Process}
The authors acknowledge the use of AI-assisted tools (such as ChatGPT) for language editing and grammar refinement during manuscript preparation. No AI tool was used for generating novel content, data analysis, or drawing conclusions. All responsibility for the accuracy and integrity of the manuscript remains with the authors.

\section*{AUTHOR CONTRIBUTIONS}
\textbf{Sungyong Chung:} Conceptualization, data curation, formal analysis, investigation, methodology, software, visualization, Writing -- original draft. \textbf{Tina Radvand:} Methodology, Validation, Writing -- review \& editing. \textbf{Alireza Talebpour:} Conceptualization, funding acquisition, methodology, project administration, supervision, validation, Writing -- review \& editing.

\bibliographystyle{arXiv}
\bibliography{Manuscript}

@article{Eisert2000,
  title = {Quantum games},
  author = {Eisert, Jens and Wilkens, Martin},
  journal = {Journal of Modern Optics},
  volume = {47},
  number = {14-15},
  pages = {2543--2556},
  year = {2000},
  publisher = {Taylor \& Francis}
}

@article{Grabbe2005,
  title = {An introduction to quantum game theory},
  author = {Grabbe, J. Orlin},
  journal = {arXiv preprint quant-ph/0506219},
  year = {2005}
}

@article{Marinatto2000,
  title = {A quantum approach to static games of complete information},
  author = {Marinatto, Luca and Weber, Tullio},
  journal = {Physics Letters A},
  volume = {272},
  number = {5-6},
  pages = {291--303},
  year = {2000},
  publisher = {Elsevier}
}

@article{Khan2018,
  title = {Quantum games: a review of the history, current state, and interpretation},
  author = {Khan, Faisal Shah and Solmeyer, Neal and Balu, Radhakrishnan and Humble, Travis S.},
  journal = {arXiv preprint arXiv:1803.07919},
  year = {2018}
}

@article{Iqbal2001,
  title = {Evolutionarily stable strategies in quantum games},
  author = {Iqbal, A. and Toor, A. H.},
  journal = {Physics Letters A},
  volume = {280},
  number = {5-6},
  pages = {249--256},
  year = {2001},
  publisher = {Elsevier}
}

@article{Chung2025,
  title = {Characterizing Lane-Changing Behavior in Mixed Traffic},
  author = {Chung, Sungyong and Talebpour, Alireza and Hamdar, Samer H.},
  journal = {arXiv preprint arXiv:2512.07219},
  year = {2025}
}

@article{Kita1999,
  title = {A merging–giveway interaction model of cars in a merging section: a game theoretic analysis},
  author = {Kita, H.},
  journal = {Transportation Research Part A: Policy and Practice},
  volume = {33},
  number = {3-4},
  pages = {305--312},
  year = {1999},
  publisher = {Elsevier}
}

@article{Kita2002,
  title = {A game theoretic analysis of merging-giveway interaction: a joint estimation model},
  author = {Kita, H. and Tanimoto, K. and Fukuyama, K.},
  journal = {Transportation and Traffic Theory in the 21st Century},
  pages = {503--518},
  year = {2002},
  publisher = {Emerald Group Publishing Limited}
}

@article{Talebpour2015,
  title = {Modeling lane-changing behavior in a connected environment: A game theory approach},
  author = {Talebpour, A. and Mahmassani, H. S. and Hamdar, S. H.},
  journal = {Transportation Research Procedia},
  volume = {7},
  pages = {420--440},
  year = {2015},
  publisher = {Elsevier}
}

@article{Tanimoto2014,
  title = {Social dilemma structures hidden behind traffic flow with lane changes},
  author = {Tanimoto, J. and Kukida, S. and Hagishima, A.},
  journal = {Journal of Statistical Mechanics: Theory and Experiment},
  volume = {2014},
  number = {7},
  pages = {P07019},
  year = {2014},
  publisher = {IOP Publishing}
}

@article{Debreu1952,
  title = {A social equilibrium existence theorem},
  author = {Debreu, Gerard},
  journal = {Proceedings of the National Academy of Sciences},
  volume = {38},
  number = {10},
  pages = {886--893},
  year = {1952},
  publisher = {National Acad Sciences}
}

@article{Aumann1974,
  title = {Subjectivity and correlation in randomized strategies},
  author = {Aumann, Robert J},
  journal = {Journal of Mathematical Economics},
  volume = {1},
  number = {1},
  pages = {67--96},
  year = {1974},
  publisher = {Elsevier}
}

@inproceedings{Ettinger2021,
  title     = {Large Scale Interactive Motion Forecasting for Autonomous Driving: The Waymo Open Motion Dataset},
  author    = {
    Ettinger, Scott and
    Cheng, Shuyang and
    Caine, Benjamin and
    Liu, Chenxi and
    Zhao, Hang and
    Pradhan, Sabeek and
    Chai, Yuning and
    Sapp, Ben and
    Qi, Charles R. and
    Zhou, Yin and
    Yang, Zoey and
    Chouard, Aurélien and
    Sun, Pei and
    Ngiam, Jiquan and
    Vasudevan, Vijay and
    McCauley, Alexander and
    Shlens, Jonathon and
    Anguelov, Dragomir
  },
  booktitle = {Proceedings of the IEEE/CVF International Conference on Computer Vision (ICCV)},
  pages     = {9710--9719},
  month     = {October},
  year      = {2021}
}

@article{An2023Bezier,
  author    = {Gihyeob An and Jun Han Bae and Alireza Talebpour},
  title     = {An optimized car-following behavior in response to a lane-changing vehicle: A Bézier curve-based approach},
  journal   = {IEEE Open Journal of Intelligent Transportation Systems},
  volume    = {4},
  pages     = {682--689},
  year      = {2023},
  publisher = {IEEE}
}

@article{An2023Platooning,
  author    = {Gihyeob An and Alireza Talebpour},
  title     = {Vehicle platooning for merge coordination in a connected driving environment: A hybrid ACC-DMPC approach},
  journal   = {IEEE Transactions on Intelligent Transportation Systems},
  volume    = {24},
  number    = {5},
  pages     = {5239--5248},
  year      = {2023},
  publisher = {IEEE}
}

@article{Rahmati2021LeftTurn,
  author    = {Yalda Rahmati and Mohammadreza Khajeh Hosseini and Alireza Talebpour},
  title     = {Helping automated vehicles with left-turn maneuvers: A game theory-based decision framework for conflicting maneuvers at intersections},
  journal   = {IEEE Transactions on Intelligent Transportation Systems},
  volume    = {23},
  number    = {8},
  pages     = {11877--11890},
  year      = {2021},
  publisher = {IEEE}
}

@article{Chung2024GapSetting,
  author    = {Sungyong Chung and Dongju Ka and Yongju Kim and Chungwon Lee},
  title     = {Gap setting control strategy for connected and automated vehicles in freeway lane-drop bottlenecks},
  journal   = {IET Intelligent Transport Systems},
  volume    = {18},
  number    = {12},
  pages     = {2641--2659},
  year      = {2024},
  publisher = {IET}
}

@article{meyer1999,
  title={Quantum strategies},
  author={Meyer, David A.},
  journal={Physical Review Letters},
  volume={82},
  number={5},
  pages={1052},
  year={1999},
  publisher={APS}
}

@article{nowak1992evolutionary,
  title={Evolutionary games and spatial chaos},
  author={Nowak, Martin A. and May, Robert M.},
  journal={Nature},
  volume={359},
  number={6398},
  pages={826--829},
  year={1992},
  publisher={Nature Publishing Group UK London}
}

@article{nash1950equilibrium,
  title={Equilibrium points in n-person games},
  author={Nash, John F.},
  journal={Proceedings of the National Academy of Sciences},
  volume={36},
  number={1},
  pages={48--49},
  year={1950},
  publisher={National Acad Sciences}
}

@article{McKelvey1995,
  author  = {McKelvey, Richard D. and Palfrey, Thomas R.},
  title   = {Quantal Response Equilibria for Normal Form Games},
  journal = {Games and Economic Behavior},
  volume  = {10},
  number  = {1},
  pages   = {6--38},
  year    = {1995}
}

\end{document}